# Performance of Solution-Processed Carbon Nanotube Field Effect Transistors with Graphene Electrodes

P R Yasasvi Gangavarapu, Punith Chikkahalli Lokesh, K N Bhat, A K Naik*

Centre for Nano Science and Engineering, Indian Institute of Science

Bangalore - 560012, India

*Corresponding Author: anaik@cense.iisc.ernet.in

**Abstract**

**This work evaluates the performance of carbon nanotube field effect transistors (CNTFET) using few layer graphene as the contact electrode material. We present the experimental results obtained on the barrier height at CNT – graphene junction using temperature dependent I-V measurements. The estimated barrier height in our devices for both holes and electrons is close to zero or slightly negative indicating the Ohmic contact of graphene with the valence and conduction bands of CNTs. In addition, we also report that there is no correlation between the barrier height and thickness of graphene.**

With silicon transistor technology reaching its practical limits, efforts to replace Si with new materials with better electronic transport properties have gained considerable attention. Among them carbon nanotube (CNT) and graphene based devices are considered to be the strong contenders to replace Si in future nanoelectronics because of their excellent electronic, mechanical and thermal properties[1–4]. However, graphene is a semimetal with zero bandgap in its pure form and is thus not suitable for transistor logic applications. On the other hand, CNTs have band gap which makes them a better choice for logic applications where large $I_{on}/I_{off}$ ratios are essential.

Although there have been reports on CNTFETs with exceptional performance[5,6], making CNTFET technology suitable for large scale production will not be possible without resolving various issues including mis-aligned CNTs, metallic CNTs during growth resulting in incorrect logic functionality among others[7–9]. One of the problems in electronic performance of CNTFETs has been the high barrier height of CNT – metal contact forming a Schottky junction which results in low ON currents[10,11]. It has been shown that low Schottky barrier contacts are achievable for both p-type[5] and n-type CNTFETs[12,13]. However, this involved the use of different metals for n and p-type devices e.g. Palladium for p-type and Scandium and Yttrium for n-type CNTFETs. The metals used for n-type CNTFETs are rare earth metals and may not be suitable for large scale production. On the other hand, graphene is considered to be an excellent contact material[14] for CNT devices because of the "homogeneous" carbon junction[15]. Furthermore, graphene/CNT hybrid devices serve as an excellent choice for transparent TFT and flexible electronic applications[16–18]. However, as discussed in the following paragraph, there are conflicting reports on the nature of the CNT – graphene junction.

There have been no reports on the suitability of graphene as contact material for n-type CNTFETs. Pei et al[19] fabricated p-type CNTFETs using exfoliated graphene as the electrode. The devices showed rectifying behaviour and the authors concluded that graphene may not be a suitable electrode material for CNTFETs. The authors further indicate that the Schottky nature of the contact is the result of work function difference between graphene and CNT. In addition, Kim et al[20] have experimentally studied the suitability of graphene as electrode materials and estimated the barrier height at the CNT – graphene junction to be 0.5eV. However, the devices considered in their work were of partially semiconducting in nature and the $I_{on}/I_{off}$ ratio (<10) was consequently poor. In contrast to these two experimental results, Cook et al reported a theoretical framework indicating the suitability of graphene as an electrode material for CNTFET applications[21]. Thus even though there have been some attempts to obtain information on the CNT – graphene contact, a detailed study to firm up the understanding of the nature of the CNT – graphene junction has not been reported so far.

In this work, we evaluate the performance of both p-type and n-type CNTFETs with exfoliated few layer graphene as the contact electrode material and we have extracted the true barrier heights for hole and electron transport at CNT – graphene junction by studying the electrical ($I_{ds} - V_{gs}$) characteristics of CNTFETs at different temperatures using the approach reported in[22]. Based on our detailed experimental study and analysis, we demonstrate that the true barrier height of CNT – graphene junction is negligibly small for both hole and electron transport indicating the suitability of graphene as electrode material for CMOS type circuits based on CNTFETs. Furthermore, by fabricating CNTFETs with graphene electrodes of varying thickness, we also demonstrate that the barrier height is practically independent of the number of graphene layers. The results also show that the values of barrier height for the fabricated devices are only few meV indicating the ohmic nature of CNT – graphene contact irrespective of whether the CNT is p-type or n-type.

Figure 1 shows the schematic diagram of the fabricated device. Graphene is mechanically exfoliated on to a 290nm thick $SiO_2$ thermally grown on a highly doped $P^{++}$ Si substrate and the graphene flakes are identified using an optical microscope. The thickness of these few layer graphene flakes is estimated using atomic force microscope (AFM). The metal electrodes are patterned using e-beam lithography followed by evaporation of Cr/Pd (2nm/50nm) for source – drain contacts (see Figure 1). A second layer of electron-beam lithography is used to open a small window to etch graphene flakes to realize graphene electrodes separated by a gap of 1μm as shown in figure 1. CNTs are aligned between these graphene electrodes using AC dielectrophoresis[23]. Commercially available CNTs dispersed in solution obtained from Nano Integris Inc (99% IsoNanotubes-S) are used as the source of CNT and the CNTs are aligned between the electrodes by applying 50kHz signal with $5V_{rms}$ between the electrodes.. The highly doped silicon substrate underneath $SiO_2$, is used as the back gate of the CNTFET. The devices in which CNTs are aligned between graphene electrodes without making any contact with the metal (Cr/Pd) pads are identified for probing. This has been ascertained using AFM. The device is next subjected to $Ar/H_2$ (at a flow rate of 500 sccm for both gases) annealing at 350 °C to remove unwanted residues accumulated during the fabrication. The resulting CNTFETs typically show p-type behaviour. To fabricate n-type CNTFETs various strategies such as vacuum annealing[24], chemical doping[25,26] and using rare earth metals like Scandium, Yttrium for source/drain contacts[12,13,27] have been proposed in literature. In our work, we used PECVD grown Silicon Nitride (100nm thick film deposited at 140°C) to fabricate air-stable n-type CNTFETs[28–30].

Figure 2(a) shows the AFM image of a CNTFET with graphene electrodes before depositing Silicon Nitride film using PECVD method. Figure 2(b) shows the transfer ($I_{ds} - V_{gs}$) characteristics of the device at room temperature and in vacuum (~1e-5 mbar) before and after deposition of 100nm of PECVD Silicon Nitride film. These plots clearly show the change in polarity of the ON and OFF conditions for the p-type to n-type CNTs (before and after nitride deposition). The $I_{on}/I_{off}$ ratio remained unchanged (~$10^5$) after Silicon Nitride deposition.

Since we have used Cr/Pd (2nm/50nm) to create electrical contact to graphene, it is important to study the nature of contact between Cr/Pd and graphene. Figures S1(a) and S1(b) (see Supplementary Information) show $I_{ds} - V_{ds}$ (at $V_{gs}$ = -15V) and $I_{ds} - V_{gs}$ (at $V_{ds}$ = 0.1V) characteristics at different temperatures of a multilayer graphene device with Cr/Pd as metal contact. These measurements are done in a low temperature probe station (Lake Shore station with Agilent B1500A Semiconductor device analyser) with vacuum of the order of 1e-6 mbar. The linear $I_{ds} - V_{ds}$ curve shown in figure S1(a) indicates the absence of Schottky barrier at Cr/Pd – graphene junction. Furthermore, unlike graphene – CNT junction (figure 2(b)), the Cr/Pd – graphene shows very little modulation of current with gate voltage as shown in figure

S1(b). This indicates that the barrier height at the metal – graphene junction is minimal and is therefore ignored in the subsequent analysis. Thus the barrier height estimated from the "Cr/Pd – Graphene – CNT" junction is entirely due to the barrier height at the "Graphene – CNT" junction.

Following the previous analysis[22], we have extracted barrier height assuming the thermionic emission transport model for the current in CNTs. The current in this model is given by

$$I_{ds} = AA^*T^2 e^{\frac{-\emptyset_B}{kT}} \left( e^{\frac{qV_{ds}}{kT}} - 1 \right) \quad \text{................ Eq. (1)}$$

Where $AA^*$ is the product of the contact area ($A$) and the effective Richardson's constant ($A^*$), $\emptyset_B$ is the barrier height, $k$ is the Boltzmann's constant, $q$ is the charge of an electron and $T$ is the temperature. The carrier transport is dominated by thermionic emission in the subthreshold regime and by field emission at higher gate voltages. The current in the subthreshold regime is given by $I_{th} = AA^*T^2 e^{\frac{-\emptyset_B}{kT}}$. Using this equation, slope of the $ln\left(\frac{I_{th}}{T^2}\right)$ vs $\left(\frac{1}{kT}\right)$ plot gives the value of barrier height (slope = $-\emptyset_B$). Figures 3(a) and 3(b) show $I_{ds} - V_{gs}$ and $I_{ds} - V_{ds}$ plots for a CNTFET with graphene (thickness of graphene is ~3.6nm) as contact material at different temperatures respectively. The $I_{ds} - V_{gs}$ data at different temperatures can be used to generate an Arrhenius type plot ($ln\left(\frac{I_{th}}{T^2}\right)$ vs $\left(\frac{1}{kT}\right)$) as shown in figure 3(c). The slopes of these plots are then used to extract the barrier height values corresponding to the different gate voltages as shown in figure 3(d). For gate voltages within the flat band voltage $V_{FB}$ ($V_{gs} < V_{FB}$), $\emptyset_B$ varies linearly with gate voltage as given in equation (2)[22,31].

$$\emptyset_B = \emptyset_{B0} + \alpha\gamma(V_{gs} - V_{FB}) \quad \text{................ Eq. (2)}$$

where $\phi_{B0}$ is the true barrier height at the metal – semiconductor junction, $\gamma = 1 + \frac{C_{IT}}{C_{ox}}$, $C_{IT}$ is the interface – trap capacitance, $C_{ox}$ is the oxide capacitance and $\alpha = 1$ for p-type devices and $\alpha = -1$ for n-type devices.

The gate voltage at which $\phi_B$ deviates from linear behaviour indicates the flat band voltage and the corresponding barrier height is the true barrier height $\phi_{B0}$ at the CNT – graphene junction. Typical true barrier heights observed in 15 distinct devices (7 p-type and 8 n-type) fabricated using the method described above are in the range of -110meV to 40meV for hole transport ($\phi_{B0-p}$) and -80meV to 65meV for electron transport ($\phi_{B0-n}$) respectively. These values are much lower compared to the value of 0.5eV[20] reported recently. Figures S2(a) – S2(d) (see Supplementary Information) show the corresponding data of an n-type CNTFET with graphene electrodes (graphene thickness = ~4.3nm).

In order to examine the effect of graphene thickness, we have fabricated CNTFETs with graphene electrodes having different thickness values and extracted true barrier heights for both holes and electrons and the results obtained are presented in figure 4. In all the devices the true barrier height is close to zero and is practically independent of graphene thickness.

The low barrier height values observed is likely the result of similar work functions of graphene and carbon nanotubes (~4.66eV) as proposed by Cook et al[21]. In addition, there have been reports where barrier heights for holes ($\phi_p$) were estimated using equation (3) under the assumption that there is no fermi level pinning at the interface[20,32].

$$\phi_p = \phi_{CNT} + \frac{E_g}{2} - \phi_m \qquad \text{................ Eq. (3)}$$

Where $\phi_p$ is barrier height for holes, $\phi_{CNT}$ is the work function of CNTs, $E_g$ is the bandgap of CNTs given by $\frac{0.78}{d(nm)}$, $d$ is diameter of CNTs (~1.4nm in this work) and $\phi_m$ is the work function of contact material (in this case graphene). The work functions of CNT and graphene are close to 4.5eV[33] and bandgap of CNTs is ~0.55eV. Thus, barrier height at CNT – graphene junction is expected to be ~0.275eV which is higher than what we have observed experimentally. However, these calculations assume that the CNTs to be defect free and undoped. In our work, we have used solution processed CNTs which are known to have defects and are possibly doped. These defects/doping will affect the work function of the CNTs. E.g. doping the CNTs with nitrogen has been shown to reduce the work function from 4.5eV to 4.1eV[34], aluminum-doped zinc-oxide nanoparticle treatment have modified the work function from 4.96eV to 4.54eV[35] and from 4.8eV to 2.4eV if Cs atom are used for intercalation on single walled CNTs[36]. Even moderate reduction of ~250meV in the work function of CNTs due to doping or defects will result in significantly lower barrier height values as observed in our devices compared to the value of 0.275eV estimated above.

**Conclusion:**

We have fabricated and evaluated the performance of both n and p-type CNTFETs with graphene as contact electrode material. We have experimentally validated that the "true barrier height" at graphene – CNT junction is close to zero for both electron and hole transport thus demonstrating the suitability of graphene as an Ohmic contact material for CNTFETs. The detailed study using different thickness of graphene indicates that the true barrier height is very low in all the cases and it is practically independent of graphene thickness. Combined with recent efforts to improve alignment in solution processed CNTs[37] and dramatic improvements in conductance[38], graphene contacts hold potential for high performance n-type and p-type CNTFETs that can match the performance of CVD based CNTFETs[39]. This work opens up the possibility of using few layer graphene as electrode material for CNT based CMOS logic circuits[9,24] and may also find interesting applications in flexible electronics and sensing applications[17].

**Supplementary Material**

Please see supplementary material for Raman characterization of the CNTs used in our work, temperature dependent I-V characteristics of n-type CNTFET and few layered graphene FET.

**Acknowledgements**

The authors would like to acknowledge the usage of National Nano-fabrication Centre (NNFC) and Micro and Nano Characterization Facility (MNCF) at CeNSE, IISc, Bangalore. The authors would like to thank Mr. M. Aamir, Dr. T. P. Sai, Dr. Shishir Kumar for helping with the Ar/$H_2$ annealing and Dr. K. L. Ganapathi for useful discussions.

**Figures**

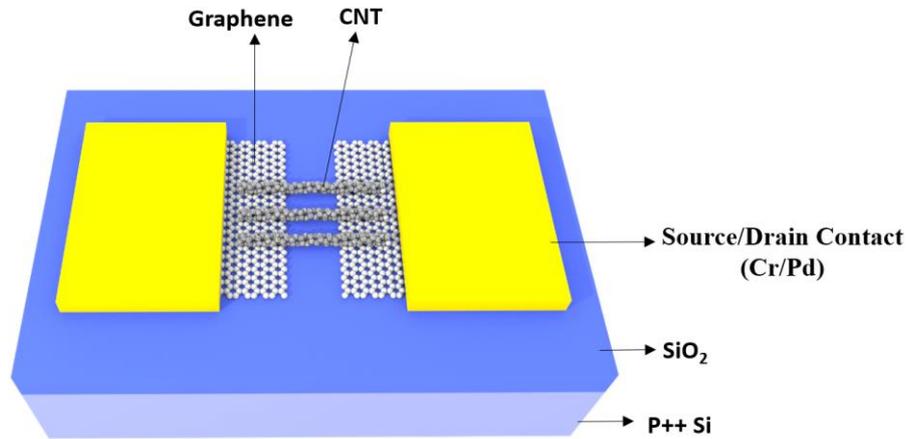

*FIG. 1. Schematic diagram of the device with CNT as the channel material and graphene as electrode material. Cr/Pd (2/50nm) metal stack is used to contact graphene for electrical measurements. The CNTs are aligned using AC dielectrophoresis such that they make contact only with graphene electrodes but not the metal. This is achieved using long (few μm) graphene flakes and CNT solution with low concentration (~20ng/ml).*

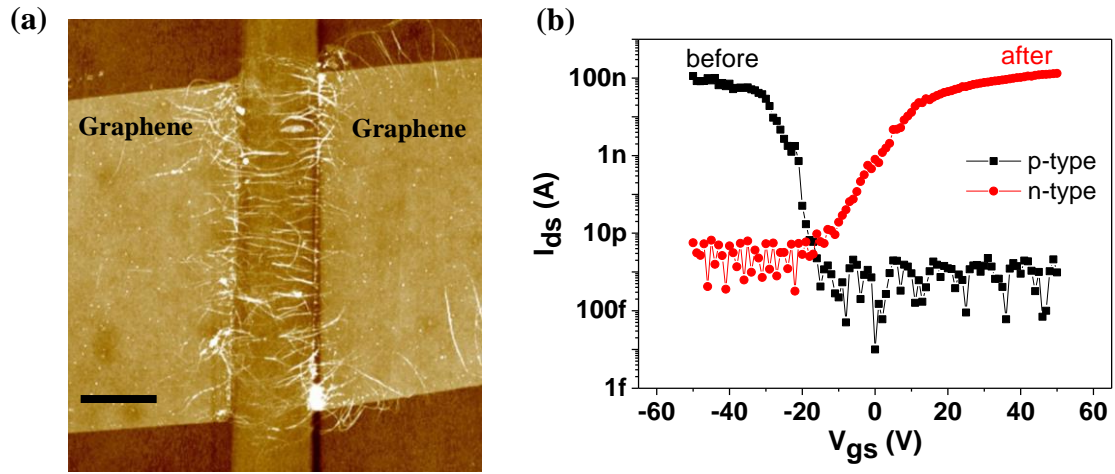

*FIG. 2.* (a) AFM image of a CNTFET with graphene contacts (without the PECVD Silicon Nitride film). Width of graphene electrodes is ~4.5μm and thickness is ~2.3nm. The CNTs have an average diameter of ~1.4nm and length (L) of the device is 1μm. Scale bar shown in the figure is 1μm. (b) $I_{ds} - V_{gs}$ characteristics of the device at $V_{ds} = 1V$, before and after Silicon Nitride deposition.

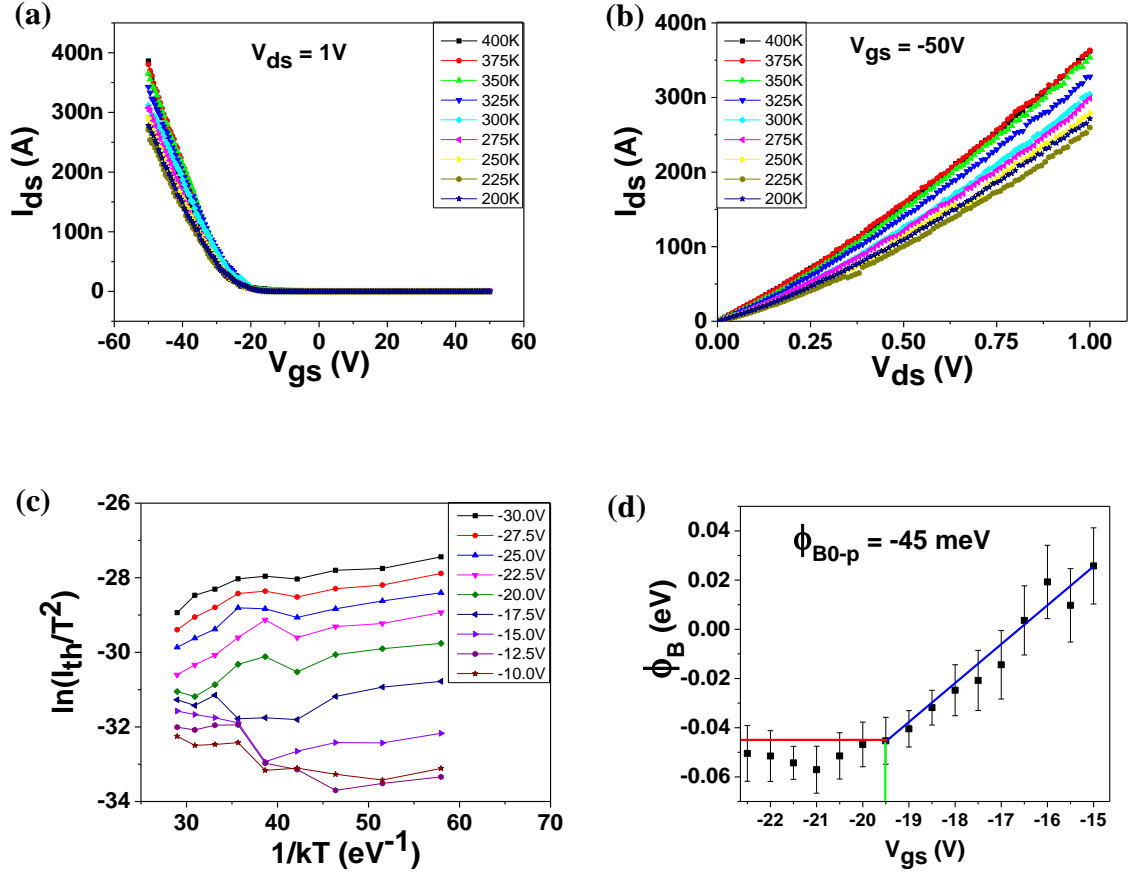

**FIG. 3.** (a) $I_{ds} - V_{gs}$ characteristics of the p-type CNTFET at different temperatures with $V_{ds} =$ 1V. The threshold voltage ($V_{th}$) of the device is -25V. (b) $I_{ds} - V_{ds}$ characteristics of the device at different temperatures with $V_{gs} = $ -50V. (c) Arrhenius plots at different gate voltages in subthreshold regime ($V_{gs} < V_{th}$) where thermionic emission is the dominant transport mechanism. (d) Barrier height ($\phi_B$) vs $V_{gs}$ plot at $V_{ds} = 1V$. The true barrier height ($\phi_{B0}$) is -45±9.5meV indicating the Ohmic nature of graphene – CNT junction. The flat band voltage ($V_{FB}$) is -19.5V.

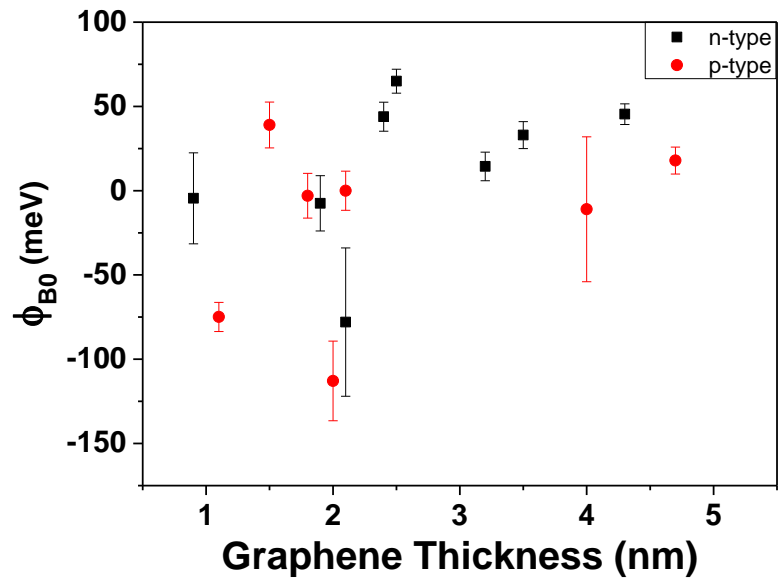

***FIG. 4.*** *Variation of $\phi_{B0}$ with graphene thickness. The values of $\phi_{B0}$ are scattered around zero and do not show any correlation with graphene thickness.*

# Supplemental Material:

# Performance of Solution-Processed Carbon Nanotube Field Effect Transistors with Graphene Electrodes

P R Yasasvi Gangavarapu, Punith Chikkahalli Lokesh, K N Bhat, A K Naik*

Centre for Nano Science and Engineering, Indian Institute of Science, Bangalore, 560012, India

**A. I-V characteristics of a graphene device:**

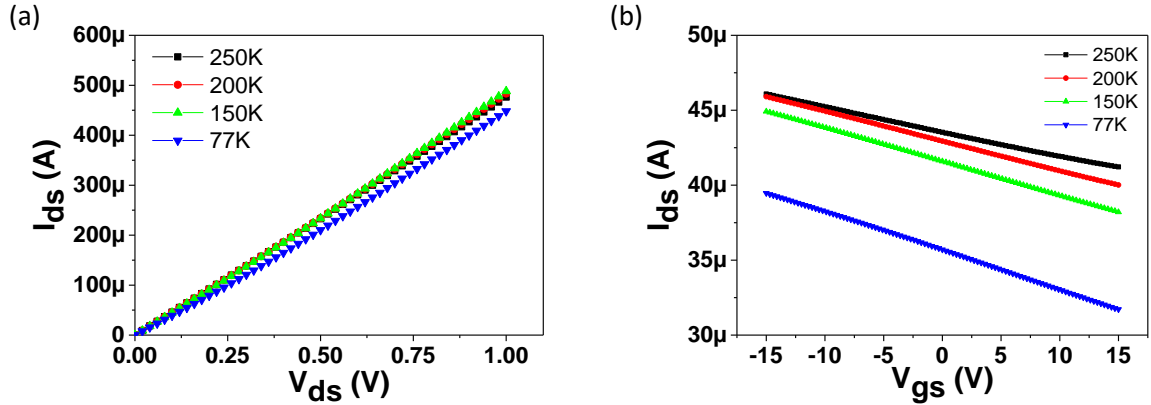

**FIG. S1.** I – V characteristics of a multilayer graphene device with Cr/Pd as the metal contact (a) $I_{ds} - V_{ds}$ characteristics of the device at $V_{gs}$ = -15V at different temperatures show almost linear response indicating zero or negligible Schottky barrier (b) $I_{ds} - V_{gs}$ characteristics of the device at $V_{ds}$ = 0.1V at different temperatures. The modulation in current with gate voltage is minimal compared to CNTFETs (see figures 2(b) and S1(b)).

## B. Barrier height calculations of an n-type CNTFET:

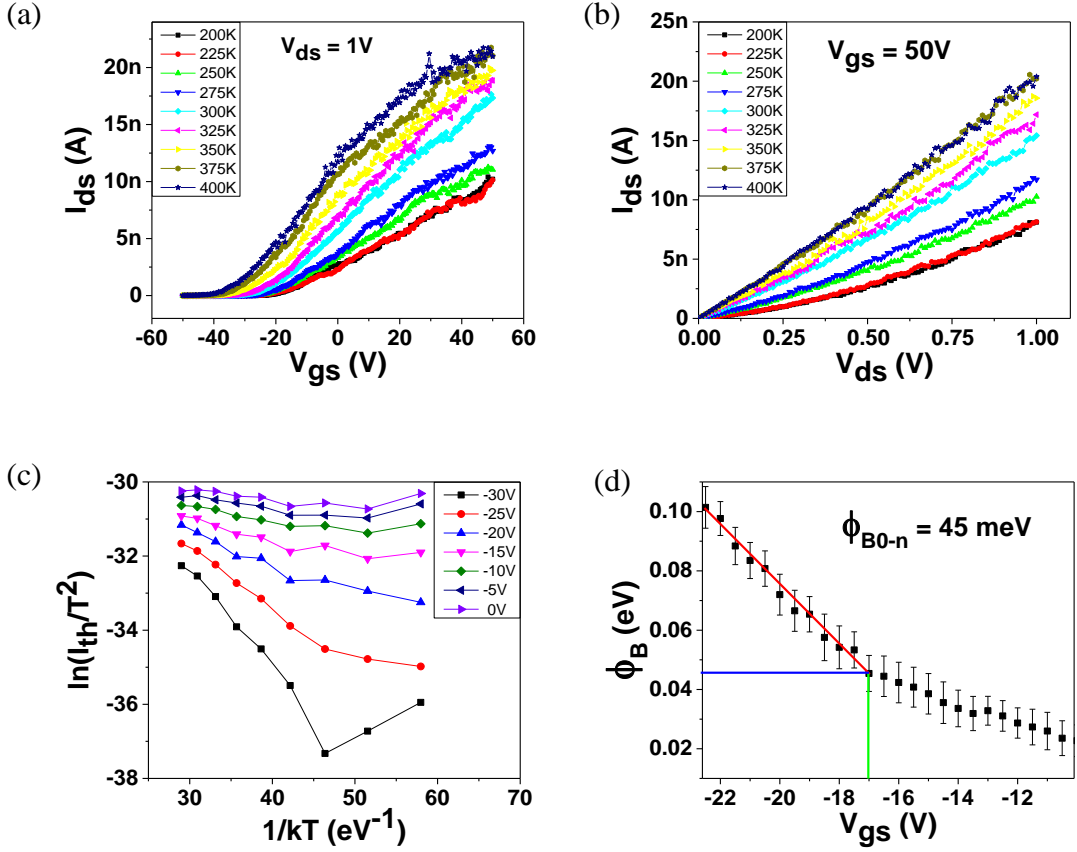

**FIG. S2.** (a) $I_{ds} - V_{gs}$ characteristics of the device at different temperatures with $V_{ds}$ = 1V. (b) $I_{ds} - V_{ds}$ characteristics of the device at different temperatures with $V_{gs}$ = -50V. (c) Arrhenius plots at different $V_{gs}$ in subthreshold regime ($V_{gs} < V_{th}$) where thermionic emission is the dominant transport mechanism. (d) Barrier height ($\phi_B$) vs $V_{gs}$ plot at $V_{ds}$ = 1V. The true barrier height for electron transport ($\phi_{B0-n}$) is 45±6meV indicating the Ohmic nature of graphene – CNT junction. The flat band voltage ($V_{FB}$) is -17V.

## C. Raman Analysis:

Raman spectra was acquired for the CNT samples (Iso-nanotubes (S) 99% from Nanointegris Inc) used in this work. Figure S5 (a) shows Raman spectra indicating D (1343cm$^{-1}$), G- (1571cm$^{-1}$), G+ (1592cm$^{-1}$) and 2D (2685cm$^{-1}$) peaks. Figure S5 (b) shows the RBM mode with the peak at ~176cm$^{-1}$ from which the diameter (d) of CNTs is estimated to be ~1.4nm (d = 238/RBM).

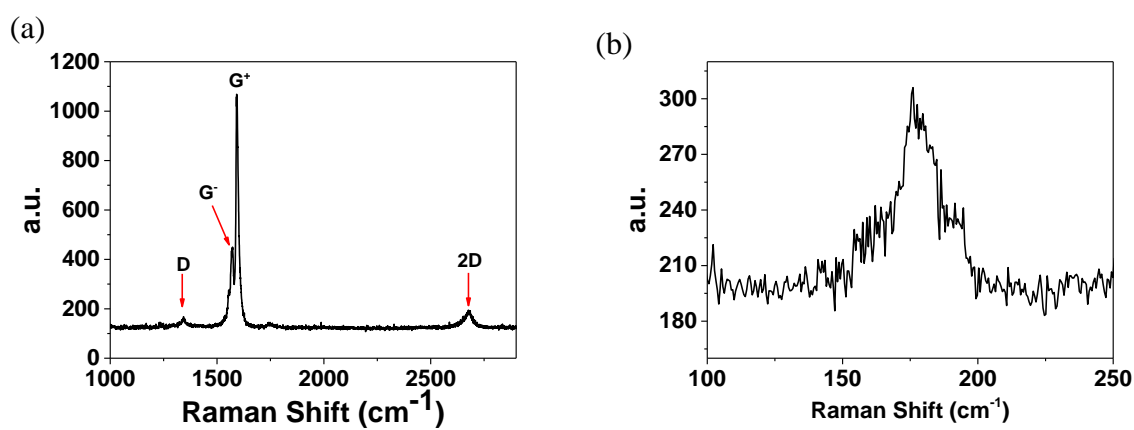

**FIG. S3.** Raman spectra of CNTs used in our work with (a) and (b) showing various peaks which are characteristic features of semiconducting CNTs.